\documentclass[american,prb,twocolumn,superscriptaddress,showpacs,twoside,letterpaper]{revtex4}
\usepackage[T1]{fontenc}
\usepackage[utf8]{inputenc}
\usepackage{amsmath}
\usepackage{graphicx}
\usepackage{esint}

\makeatletter

\providecommand{\tabularnewline}{\\}

\@ifundefined{textcolor}{}
{%
 \definecolor{BLACK}{gray}{0}
 \definecolor{WHITE}{gray}{1}
 \definecolor{RED}{rgb}{1,0,0}
 \definecolor{GREEN}{rgb}{0,1,0}
 \definecolor{BLUE}{rgb}{0,0,1}
 \definecolor{CYAN}{cmyk}{1,0,0,0}
 \definecolor{MAGENTA}{cmyk}{0,1,0,0}
 \definecolor{YELLOW}{cmyk}{0,0,1,0}
}

\makeatother

\usepackage{babel}
\begin{document}

\title{Ab-initio study of the thermopower of biphenyl-based single-molecule
junctions}

\author{M. B\"urkle}

\email{marius.buerkle@kit.edu}

\affiliation{Institute of Theoretical Solid State Physics, Karlsruhe Institute
of Technology, D-76131 Karlsruhe, Germany}

\affiliation{DFG Center for Functional Nanostructures, Karlsruhe Institute of
Technology, D-76131 Karlsruhe, Germany}

\author{L. A. Zotti}

\affiliation{Departamento de F\'isica Te\'orica de la Materia Condensada, Universidad
Aut\'onoma de Madrid, E-28049 Madrid, Spain}

\author{J. K. Viljas}

\affiliation{Low Temperature Laboratory, Aalto University, P.O.\ Box 15100, FIN-00076
Aalto, Finland}

\affiliation{Department of Physics, P.O.\ Box 3000, FIN-90014 University of Oulu,
Finland}

\author{D. Vonlanthen}

\affiliation{Department of Chemistry, University of Basel, CH-4056 Basel, Switzerland}

\author{A. Mishchenko}

\affiliation{Department of Chemistry and Biochemistry, University of Bern, CH-3012
Bern, Switzerland}

\author{T. Wandlowski}

\affiliation{Department of Chemistry and Biochemistry, University of Bern, CH-3012
Bern, Switzerland}

\author{M. Mayor}

\affiliation{DFG Center for Functional Nanostructures, Karlsruhe Institute of
Technology, D-76131 Karlsruhe, Germany}

\affiliation{Department of Chemistry, University of Basel, CH-4056 Basel, Switzerland}

\affiliation{Institute for Nanotechnology, Karlsruhe Institute of Technology,
D-76344 Eggenstein-Leopoldshafen, Germany}

\author{G. Sch\"on}

\affiliation{Institute of Theoretical Solid State Physics, Karlsruhe Institute
of Technology, D-76131 Karlsruhe, Germany}

\affiliation{DFG Center for Functional Nanostructures, Karlsruhe Institute of
Technology, D-76131 Karlsruhe, Germany}

\affiliation{Institute for Nanotechnology, Karlsruhe Institute of Technology,
D-76344 Eggenstein-Leopoldshafen, Germany}

\author{F. Pauly}

\affiliation{Institute of Theoretical Solid State Physics, Karlsruhe Institute
of Technology, D-76131 Karlsruhe, Germany}

\affiliation{DFG Center for Functional Nanostructures, Karlsruhe Institute of
Technology, D-76131 Karlsruhe, Germany}

\affiliation{Molecular Foundry, Lawrence Berkeley National Laboratory, Berkeley,
California 94720, USA}

\pacs{85.65.+h, 85.80.Fi, 73.63.Rt, 81.07.Pr}
\begin{abstract}
Employing ab-initio electronic structure calculations combined with
the non-equilibrium Green's function technique we study the dependence
of the thermopower $Q$ on the conformation in biphenyl-based single-molecule
junctions. For the series of experimentally available biphenyl molecules,
alkyl side chains allow us to gradually adjust the torsion angle $\varphi$
between the two phenyl rings from $0^{\circ}$ to $90^{\circ}$ and
to control in this way the degree of $\pi$-electron conjugation.
Studying different anchoring groups and binding positions, our theory
predicts that the absolute values of the thermopower decrease slightly
towards larger torsion angles, following an $a+b\cos^{2}\varphi$
dependence. The anchoring group determines the sign of $Q$ and $a,b$,
simultaneously. Sulfur and amine groups give rise to $Q,a,b>0$, while
for cyano $Q,a,b<0$. The different binding positions can lead to
substantial variations of the thermopower mostly due to changes in
the alignment of the frontier molecular orbital levels and the Fermi
energy. We explain our ab-initio results in terms of a $\pi$-orbital
tight-binding model and a minimal two-level model, which describes
the pair of hybridizing frontier orbital states on the two phenyl
rings. The variations of the thermopower with $\varphi$ seem to be
within experimental resolution.
\end{abstract}
\maketitle

\section{Introduction}

Tailored nanostructures hold promise for improved efficiencies of
thermoelectric materials.\cite{Vineis2010,Dubi2011,Malen2010} For
this reason there is a growing interest to gain a better understanding
of the role of interfaces on thermoelectric properties at the atomic
scale. Controlled metal-organic interfaces can be studied using single-molecule
junctions, and recently the thermopower of these systems was determined
in first experiments.\cite{Reddy2007} While the thermopower (or Seebeck
coefficient) of metallic atomic contacts was measured already several
years ago,\cite{Ludoph1999} molecular junctions offer fascinating
possibilities to adjust thermoelectric properties due to the control
over chemical synthesis and interface structure. Ref.~\onlinecite{Reddy2007}
and subsequent experimental studies thus explored the influence of
different parameters on the thermopower, such as molecule length,\cite{Reddy2007,Malen2009a,Tan2011}
substituents,\cite{Baheti2008} anchoring groups,\cite{Baheti2008,Tan2011,Widawsky2011}
or electrode metal.\cite{Yee2011}

On the theory side, the electronic contribution to the thermopower
explains important experimental observations.\cite{Paulsson2003}
We have shown recently that the thermopower of metallic atomic contacts,
which serve as reference systems in molecular electronics, can be
understood by considering the electronic structure of disordered junction
geometries.\cite{Pauly2011} Using molecular dynamics simulations
of many junction stretching processes combined with tight-binding-based
electronic structure and transport calculations, we found thermopower-conductance
scatter plots similar to the low-temperature experiment.\cite{Ludoph1999}
Such a statistical analysis, although highly desirable for molecular
junctions, is complicated by the time-consuming electronic structure
calculations needed to describe these heteroatomic systems. Still,
early studies of the thermopower based on density functional theory
(DFT) for selected geometries explained crucial trends, such as the
dependence of the thermopower on molecule length\cite{Pauly2008}
or the influence of substituents and anchoring groups.\cite{Pauly2008,Ke2009}
Since the experiments on the thermopower of molecular junctions were
all performed at room temperature until now, finite temperature effects
may play a role. They can impact the thermopower by fluctuations of
the junction geometry and electron-vibration couplings.\cite{Galperin2008,Sergueev2011,Pauly2011}
While the quantification constitutes an interesting challenge for
future work, we will focus here on the purely electronic effects in
static ground-state contact structures. 

\begin{figure}[b]
\centering{}\includegraphics[width=0.7\columnwidth]{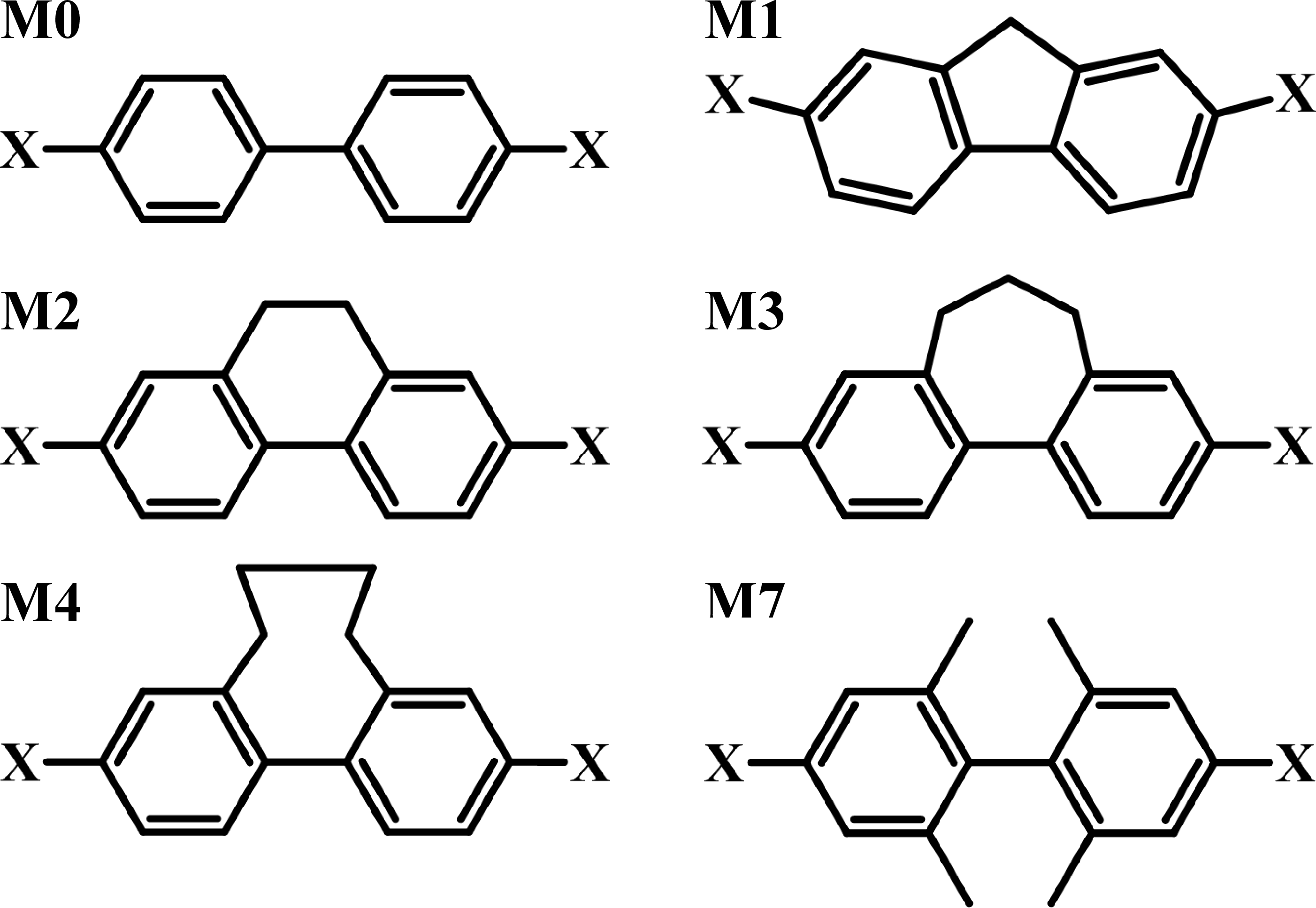}\caption{\label{fig:Chemical-structure}Chemical structure of the studied biphenyl
molecules with X standing either for the S, $\mbox{NH}_{2}$, or CN
anchoring group.}
\end{figure}

\begin{figure*}
\centering{}\includegraphics[width=0.9\linewidth]{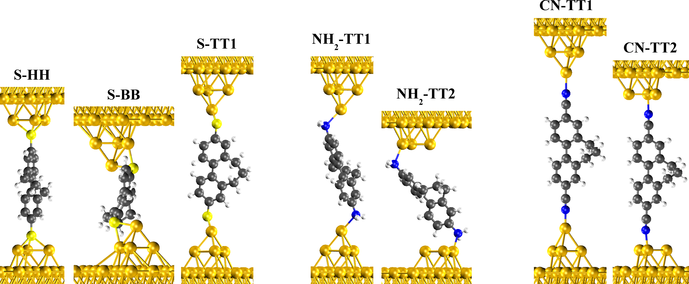}\caption{\label{fig:geos}(Color online) Analyzed types of junctions, shown
for M3. For S anchors we consider hollow, bridge, and top binding
sites to Au with the corresponding contact geometries called S-HH,
S-BB, and S-TT1, respectively. For $\mbox{NH}_{2}$ and CN we consider
binding to single Au atoms in two different top positions with the
contacts named $\mbox{NH}_{2}$-TT1, $\mbox{NH}_{2}$-TT2 and CN-TT1,
CN-TT2. }
\end{figure*}

An interesting aspect, not yet addressed in the experiments is the
influence of conjugation on the thermopower $Q$. For the conductance,
such studies were carried out by different groups with biphenyl molecules.\cite{Venkataraman2006,Mishchenko2010,Mishchenko2011}
The torsion angle $\varphi$ between the phenyl rings was adjusted
stepwise by use of appropriate side groups. While such substituents
may have a parasitic shifting effect on energies of current-carrying
molecular orbitals, the changes in conformation, which control the
degree of $\pi$-electron conjugation, turned out to dominate the
behavior of the conductance.\cite{Venkataraman2006} The systematic
series of biphenyl molecules of Refs.~\onlinecite{Vonlanthen2009,Mishchenko2010,Rotzler2010,Mishchenko2011}
uses alkyl chains of various lengths and methyl groups to adjust $\varphi$
and avoids strongly electron-donating and electron-withdrawing substituents.
Hence it seems ideal for determining the influence of conjugation
on thermopower.

Theoretical work has considered the behavior of $Q$ when $\varphi$
is changed continuously for the thiolated biphenyl molecule contacted
to gold (Au) electrodes.\cite{Pauly2008,Finch2009} Both studies agree
in the fact that $Q$ is positive for all $\varphi$. However, while
we predicted $Q$ to decrease with increasing $\varphi$ based on
DFT calculations and a $\pi$-orbital tight-binding model (TBM),\cite{Pauly2008}
work of Finch \textit{et al.}\cite{Finch2009} suggested the opposite
for this idealized system. In this study we clarify this contradiction
and demonstrate with the help of a two-level model (2LM) that for
the off-resonant transport situation the absolute value of $Q$ is
expected to decrease when the molecule changes from planar to perpendicular
ring orientation. This confirms our previous conclusions. More importantly,
this work explores the possibility to measure the dependence of $Q$
on $\varphi$ for the experimentally relevant family of molecules
presented in Refs.~\onlinecite{Vonlanthen2009,Mishchenko2010,Rotzler2010,Mishchenko2011}.

Using DFT calculations of the electronic structure combined with the
Landauer-B\"uttiker scattering formalism expressed with Green's function
techniques, we determine the thermopower of biphenyl-derived molecules
connected to gold electrodes. The molecules investigated are displayed
in Fig.~\ref{fig:Chemical-structure}. Alkyl chains, one to four
CH$_{2}$ units long, allow to change $\varphi$ gradually from $0^{\circ}$
to $60^{\circ}$. To achieve $\varphi\approx90^{\circ}$, we included
in addition M7, and as a reference also M0, the ``standard'' biphenyl
molecule. For each of the molecules in Fig.~\ref{fig:Chemical-structure}
we will explore the three different anchoring groups sulfur (S), amine
(NH$_{2}$), and cyano (CN) in various binding geometries.

This work is organized as follows. In Sec.~\ref{sec:Theoretical-methods}
we introduce the theoretical procedures used in this work. Sec.~\ref{sec:Results-and-Discussion}
presents the main results. We start by discussing models to describe
the $\varphi$ dependence of the thermopower, show the DFT-based results
for $Q$, and provide further insights by discussing their relation
to the predictions of the TBM and the 2LM. The paper ends with the
conclusions in Sec.~\ref{sec:Conclusions}.

\begin{table*}
\begin{centering}
\begin{tabular}{c||r@{\extracolsep{0pt}.}lr@{\extracolsep{0pt}.}l|r@{\extracolsep{0pt}.}lr@{\extracolsep{0pt}.}l|r@{\extracolsep{0pt}.}lr@{\extracolsep{0pt}.}l||r@{\extracolsep{0pt}.}lr@{\extracolsep{0pt}.}l|r@{\extracolsep{0pt}.}lr@{\extracolsep{0pt}.}l||r@{\extracolsep{0pt}.}lr@{\extracolsep{0pt}.}l|r@{\extracolsep{0pt}.}lr@{\extracolsep{0pt}.}l}
 & \multicolumn{4}{c|}{S-HH} & \multicolumn{4}{c|}{S-BB} & \multicolumn{4}{c||}{S-TT1} & \multicolumn{4}{c|}{NH$_{2}$-TT1} & \multicolumn{4}{c||}{NH$_{2}$-TT2} & \multicolumn{4}{c|}{CN-TT1} & \multicolumn{4}{c}{CN-TT2}\tabularnewline
 & \multicolumn{2}{c}{$\varphi$} & \multicolumn{2}{c|}{$Q$} & \multicolumn{2}{c}{$\varphi$} & \multicolumn{2}{c|}{$Q$} & \multicolumn{2}{c}{$\varphi$} & \multicolumn{2}{c||}{$Q$} & \multicolumn{2}{c}{$\varphi$} & \multicolumn{2}{c|}{$Q$} & \multicolumn{2}{c}{$\varphi$} & \multicolumn{2}{c||}{$Q$} & \multicolumn{2}{c}{$\varphi$} & \multicolumn{2}{c|}{$Q$} & \multicolumn{2}{c}{$\varphi$} & \multicolumn{2}{c}{$Q$}\tabularnewline
\hline 
\hline 
M0 & 35&1 & 0&002 & 12&9 & 1&140 & 17&9 & 0&907 & 32&9 & 0&343 & 33&2 & 0&150 & 33&1 & -2&389 & 35&3 & -1&566\tabularnewline
M1 & 0&3 & 0&064 & 0&1 & 1&313 & 0&4 & 1&280 & 0&2 & 0&436 & 0&7 & 0&201 & 0&7 & -2&281 & 0&1 & -1&423\tabularnewline
M2 & 19&8 & 0&048 & 19&3 & 1&208 & 17&2 & 1&266 & 20&4 & 0&429 & 19&1 & 0&191 & 19&5 & -2&252 & 20&2 & -1&404\tabularnewline
M3 & 42&3 & 0&032 & 42&1 & 1&127 & 42&0 & 1&200 & 46&3 & 0&382 & 46&3 & 0&189 & 45&1 & -1&878 & 46&4 & -1&158\tabularnewline
M4 & 60&7 & 0&034 & 53&0 & 1&006 & 60&8 & 0&981 & 61&6 & 0&400 & 58&4 & 0&158 & 59&2 & -1&657 & 59&9 & -0&938\tabularnewline
M7 & 89&6 & 0&019 & 84&0 & 0&298 & 83&4 & 0&981 & 87&4 & 0&191 & 87&3 & 0&091 & 89&6 & -1&173 & 89&9 & -0&632\tabularnewline
\end{tabular}
\par\end{centering}

\centering{}\caption{\label{tab:phiQtable-DFT}Torsion angle {\small $\varphi$} in units
of degrees and the thermopower {\small $Q$} at $T=10\,\mbox{K}$
in units of $\mu$V/K for all junction geometries.}
\end{table*}

\section{Theoretical methods\label{sec:Theoretical-methods}}

\subsection{Electronic structure and contact geometries}

We determine the electronic structure and contact geometries in the
framework of DFT. All our calculations are performed with the quantum
chemistry package TURBOMOLE 6.3,\cite{Ahlrichs1989} and we use the
gradient-corrected BP86 exchange-correlation functional.\cite{Perdew1986,Becke1988}
For the basis set, we employ def2-SV(P) which is of split-valence
quality with polarization functions on all non-hydrogen atoms.\cite{Weigend2005}
For Au an effective core potential efficiently deals with the innermost
60 electrons,\cite{Andrae1990} while the basis set provides an all-electron
description for the rest of the atoms in this work.

The contact geometries for the S-terminated molecules are those of
Ref.~\onlinecite{Burkle2012}. For the NH$_{2}$ and CN anchors we
proceed as described in Refs.~\onlinecite{Mishchenko2011,Burkle2012}
and use for consistency the electrode geometry from Ref.~\onlinecite{Burkle2012}.

\subsection{Charge transport}

We determine charge transport properties within the Landauer-B\"uttiker
formalism. The transmission function $\tau(E)$, describing the energy-dependent
transmission probability of electrons through the nanostructure, is
calculated with non-equilibrium Green's function techniques. The Green's
functions are constructed by use of the DFT electronic structure as
obtained for the ground-state molecular junction geometries. A detailed
description of our quantum transport method is given in Ref.~\onlinecite{Pauly2008a}.

The thermopower at the average temperature $T$ is defined as the
ratio of the induced voltage difference $\Delta V$ in the steady
state and the applied temperature difference $\Delta T$ between the
ends of a sample, $Q(T)=\left.-\left(\Delta V/\Delta T\right)\right|_{I=0}$.
In the Landauer-B\"uttiker formalism the electronic contribution
to the thermopower can be expressed as\cite{Houten1992}

\begin{equation}
Q(T)=-\dfrac{K_{1}(T)}{eTK_{0}(T)}\label{eq:Qfull}
\end{equation}
with $K_{n}(T)=\int dE\tau(E)(E-\mu)^{n}[-\partial f(E,T)/\partial E]$,
the absolute value of the electron charge $e=\left|e\right|$, the
Fermi function $f(E,T)=\{\exp[(E-\mu)/k_{B}T]+1\}^{-1}$, the Boltzmann
constant $k_{B}$, and the chemical potential $\mu\approx E_{F}=-5$
eV, which approximately equals the Fermi energy $E_{F}$ of the Au
electrodes. At low temperatures, Eq.~(\ref{eq:Qfull}) simplifies
to\cite{Paulsson2003} 

\begin{equation}
Q(T)=-q(T)\left.\frac{\partial_{E}\tau(E)}{\tau(E)}\right|_{E_{F}}.\label{eq:QlowT}
\end{equation}
with the prefactor $q(T)=\pi^{2}k_{B}^{2}T/(3e)$ depending linearly
on temperature.

In the following the thermopower is calculated, if not otherwise indicated,
for a low temperature of $T=10\,\mbox{K}$, where our theory is expected
to apply best. For the DFT-based results presented below we determine
$Q$ by means of Eq.~(\ref{eq:Qfull}), i.e.\ by taking into account
the thermal broadening of the electrodes. For the molecular junctions
studied here, the differences to the values obtained via Eq.~(\ref{eq:QlowT})
often turn out to be small even at room temperature ($T=300\,\mbox{K}$).
Hence thermopower values for higher $T$ can be estimated using the
values at $10\,\mathrm{K}$ through $Q(T)\approx(T/10\,\textrm{K})\times Q(10\,\mathrm{K})$.
Since we are not primarily interested in the temperature dependence
of $Q$ in this work, we suppress from here on the temperature argument.

\section{Results and discussion\label{sec:Results-and-Discussion}}

\subsection{Models for the angle-dependent thermopower\label{sub:Qphi-Models}}

In Ref.~\onlinecite{Pauly2008} we argued that the thermopower should
depend on the torsion angle as 
\begin{equation}
Q_{\varphi}\approx a+b\cos^{2}\varphi.\label{eq:Q_abcos2}
\end{equation}
Our argument was based on the observation that for a $\pi$-orbital
TBM in the off-resonant transport situation, the transmission of the
biphenyl molecule can be expanded in powers of $\cos^{2}\varphi$
as\cite{Pauly2008,Pauly2008b,Viljas2008} 
\begin{equation}
\tau_{\varphi}(E)=\alpha_{2}(E)\cos^{2}\varphi+\alpha_{4}(E)\cos^{4}\varphi+O(\cos^{6}\varphi).\label{eq:TE_cos2expand}
\end{equation}
Conductance measurements\cite{Venkataraman2006,Mishchenko2010,Mishchenko2011}
and corresponding DFT calculations,\cite{Pauly2008b} which both determine
the transmission at the Fermi energy, show that $\alpha_{2}$ is the
dominant term. The leading term in the $\varphi$ dependence of $Q$
is obtained from Eq.~(\ref{eq:QlowT}) by taking into account the
energy dependence of the expansion coefficients $\alpha_{j}(E)$ and
considering the terms up to $j=4$. Then, we obtain Eq.~(\ref{eq:Q_abcos2})
with 
\begin{eqnarray}
a & = & -q\left.\frac{\partial_{E}\alpha_{2}(E)}{\alpha_{2}(E)}\right|_{E=E_{F}},\label{eq:a_TE}\\
b & = & -q\left.\frac{\alpha_{2}(E)\partial_{E}\alpha_{4}(E)-\alpha_{4}(E)\partial_{E}\alpha_{2}(E)}{\alpha_{2}(E)^{2}}\right|_{E=E_{F}}.\label{eq:b_TE}
\end{eqnarray}
While this model uses minimal information about the biphenyl molecular
junction, a disadvantage is that the magnitude and energy dependence
of the coefficients $\alpha_{2}$ and $\alpha_{4}$ are a priori unknown.

An alternative strategy is to use the 2LM of Ref.~\onlinecite{Mishchenko2010}.
This minimal model explains the $\cos^{2}\varphi$ law of the conductance
by considering the pair of hybridizing frontier orbital resonances
of the phenyl rings which are closest to $E_{F}$. Within this model,
the transmission is given by\cite{Mishchenko2010} 
\begin{equation}
\tau_{\varphi}(E)=\left|\frac{\tilde{\Gamma}\tilde{t}\cos\varphi}{(E-\tilde{\varepsilon}_{s}(\varphi)-i\tilde{\Gamma}/2)(E-\tilde{\varepsilon}_{a}(\varphi)-i\tilde{\Gamma}/2)}\right|^{2}\label{eq:TE_2LM}
\end{equation}
with $\tilde{\varepsilon}_{s,a}(\varphi)=\tilde{\varepsilon}_{0}\pm\tilde{t}\cos\varphi$.
Here, $\tilde{\varepsilon}_{0}$ is the relevant frontier molecular
orbital energy of the individual phenyl ring. For the biphenyl molecule,
it can be determined as the highest occupied molecular orbital (HOMO)
or lowest unoccupied molecular orbital (LUMO) energy for vanishing
inter-ring coupling $\tilde{t}\cos\varphi$ at $\varphi=90^{\circ}$.
The angle-dependent inter-ring coupling leads to a splitting of the
pair of degenerate levels $\tilde{\varepsilon}_{0}$ at energies $\tilde{\varepsilon}_{s,a}(\varphi)=\tilde{\varepsilon}_{0}\pm\tilde{t}\cos\varphi$
with symmetric and antisymmetric wavefunctions, respectively. In addition,
we have made the wide-band approximation with a symmetric and energy-independent
coupling $\tilde{\Gamma}$ to the left and right phenyl rings. The
2LM is hence characterized by the parameters $\tilde{\varepsilon}_{0},\tilde{t},\tilde{\Gamma}$.

We set $\tilde{\varepsilon}=\tilde{\varepsilon}_{0}-E_{F}$, $\tilde{x}=\tilde{t}\cos\varphi/\sqrt{\tilde{\varepsilon}{}^{2}+\tilde{\Gamma}^{2}/4}$,
and assume $\left|\tilde{x}\right|\ll1$. Performing a Taylor expansion
in $\tilde{x}$, we obtain Eq.~(\ref{eq:Q_abcos2}) with 
\begin{eqnarray}
a & = & -q\frac{4\tilde{\varepsilon}}{\tilde{\varepsilon}^{2}+\tilde{\Gamma}^{2}/4},\label{eq:a_2LM}\\
b & = & -q\frac{4\tilde{t}^{2}\tilde{\varepsilon}\left(\tilde{\varepsilon}^{2}-3\tilde{\Gamma}^{2}/4\right)}{\left(\tilde{\varepsilon}^{2}+\tilde{\Gamma}^{2}/4\right)^{3}}.\label{eq:b_2LM}
\end{eqnarray}
These expressions predict that the sign of $a,b$ is determined by
$\tilde{\varepsilon}$. Thus, when $\tilde{\varepsilon}$ changes
sign, $a,b$ change sign at the same time. In the typical off-resonant
transport situation $\left|\tilde{\varepsilon}\right|\gg\left|\tilde{t}\right|,\tilde{\Gamma}$,
the sign of $a\approx-q4/\tilde{\varepsilon}$ and $b\approx-q4\tilde{t}^{2}/\tilde{\varepsilon}{}^{3}$
is identical. However, $b$ may be of a different sign than $a$ in
a more on-resonant case when the broadening $\tilde{\Gamma}$ is of
a similar size as $\tilde{\varepsilon}$, i.e., when $\tilde{\varepsilon}^{2}-3\tilde{\Gamma}^{2}/4$
changes sign.

\subsection{Thermopower based on density functional theory\label{sub:DFT-results}}

\begin{figure}[tb]
\centering{}\includegraphics[width=1\columnwidth]{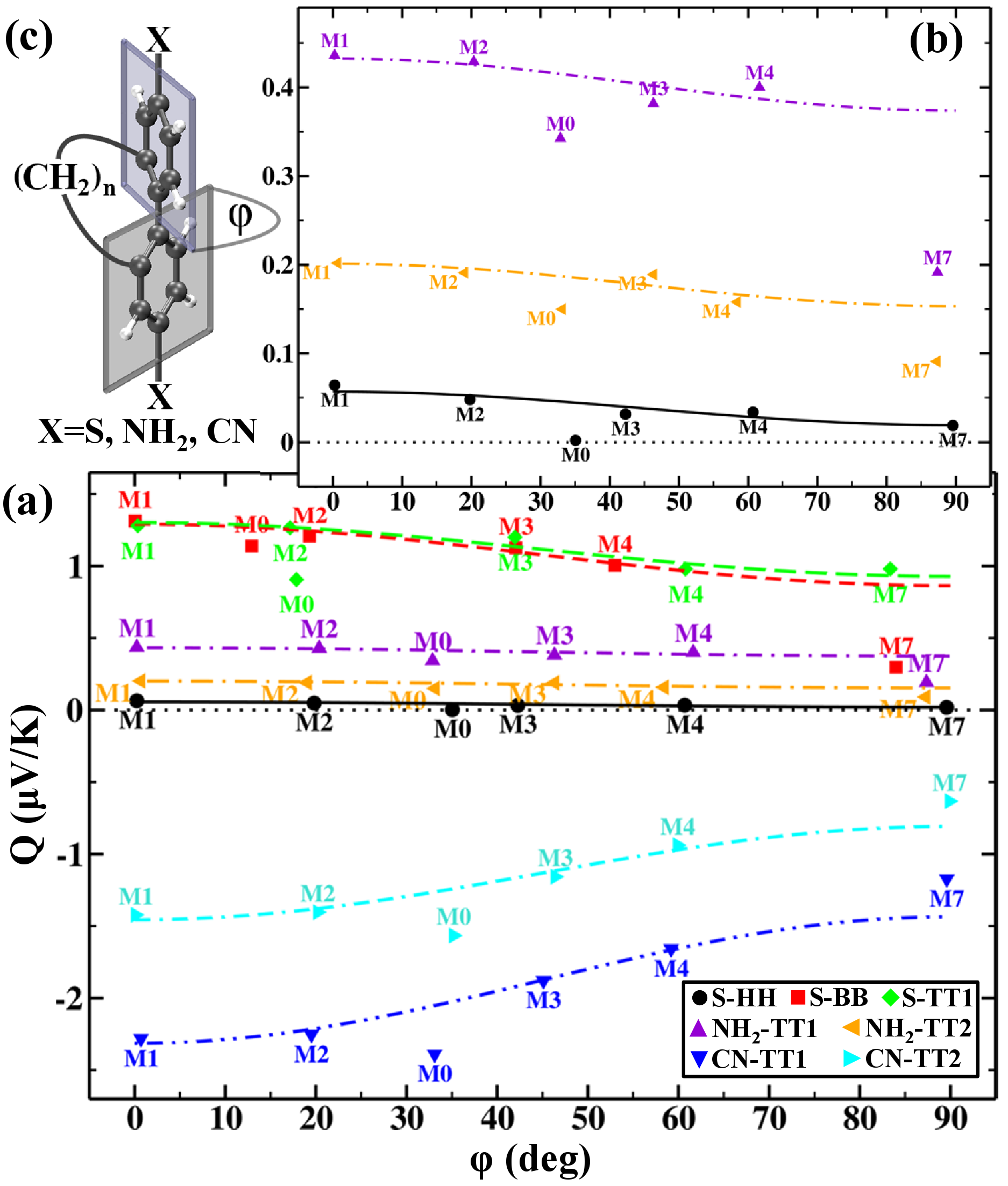}\caption{\label{fig:QphiDFT}(Color online) (a) Evolution of $Q$ with increasing
$\varphi$ for all contact geometries. The symbols represent the thermopower
values calculated with DFT at $T=10$~K, and the lines are obtained
by fitting Eq.~(\ref{eq:Q_abcos2}) to M1-M4 for each type of junction.
(b) Zoom in on the $Q$ values for S-HH, $\mbox{NH}_{2}$-TT1, and
$\mbox{NH}_{2}$-TT2. (c) Schematic of the studied biphenyl derivatives
and definition of the torsion angle $\varphi$.}
\end{figure}

For each of the biphenyl molecules in Fig.~\ref{fig:Chemical-structure},
we study the three different anchoring groups $\mathrm{X}=\mathrm{S},\mathrm{NH}_{2},\mathrm{CN}$
and select a total of seven contact geometries, as displayed in Fig.~\ref{fig:geos}.
For S anchors we choose three representative binding sites,\cite{Yu2006,Burkle2012}
where S binds covalently either to three Au atoms in the hollow position
(S-HH), to two of them in the bridge position (S-BB), or to a single
one in the top position (S-TT1). $\mbox{NH}_{2}$- and CN-terminated
molecules bind selectively to a single Au electrode atom at each side
via the nitrogen lone pair.\cite{Venkataraman2006a,Mishchenko2011}
Thus, we consider two different top sites for $\mbox{NH}_{2}$ ($\mbox{NH}_{2}$-TT1,
$\mbox{NH}_{2}$-TT2) and CN (CN-TT1, CN-TT2), respectively. 

In Table \ref{tab:phiQtable-DFT} we summarize the torsion angle $\varphi$,
which is defined as the dihedral angle between the two phenyl rings
(see Fig.~\ref{fig:QphiDFT}), and the thermopower for all 42 molecular
junctions studied. The data is presented graphically in Fig.~\ref{fig:QphiDFT}
by plotting $Q$ as a function of $\varphi$ for each of the seven
types of junctions in Fig.~\ref{fig:geos}. We notice that the sign
of the thermopower is determined by the anchoring group. For the electron-donating
S and NH$_{2}$ linkers\cite{Hansch1991} the energy of the $\pi$-electron
system of the molecules is increased compared to the hydrogen-terminated
case ($\mathrm{X}=\mathrm{H}$ in Fig.~\ref{fig:Chemical-structure}).
The HOMO energy is therefore close to $E_{F}$, as visible also from
the transmission curves in Fig.~\ref{fig:DFT-TB-2LM}. The hole conduction
through the HOMO yields $Q>0$, in agreement with previous experimental\cite{Reddy2007,Malen2009a}
and theoretical results.\cite{Pauly2008,Ke2009} In contrast to this,
for the electron-withdrawing CN anchoring group\cite{Hansch1991}
we have electron transport through the LUMO\cite{Baheti2008,Ke2009,Zotti2010,Mishchenko2011}
(see also Fig.~\ref{fig:DFT-TB-2LM}), and consequently $Q<0$.

Considering the absolute values of the thermopower, Fig.~\ref{fig:QphiDFT}
shows that $Q$ can differ markedly for the types of contact geometries.
Given the off-resonant transport situation suggested by the transmission
curves in Fig.~\ref{fig:DFT-TB-2LM} and using Eqs.~(\ref{eq:a_2LM})
and (\ref{eq:b_2LM}), we can understand the results by changes in
the level alignment $\tilde{\varepsilon}$. As we will discuss in
more detail below in Sec.~\ref{sub:Ana-TBM-2LM}, level broadenings
$\tilde{\Gamma}$ and couplings $\tilde{t}$ play no important role
in that respect. The level alignment is determined by the charge transfer
between the molecule and the electrodes, which is sensitive to the
binding site of the anchoring group at the molecule-metal interface.
For the thiolated molecules we find that the thermopower for S-BB
and S-TT1 is comparable, but the values are significantly larger than
those for S-HH. This behavior is related to our recent findings for
the conductance of the thiolated molecules, where top and bridge geometries
yield similar but much larger conductances than those with hollow
sites.\cite{Burkle2012} Both observations are due to a HOMO level
which is more distant from $E_{F}$ for S-HH as compared to S-BB and
S-TT1. We explain this by the leakage of electrons from the molecule,
including the S atoms, to the Au electrodes, when going from the S-TT1
over the S-BB to the S-HH geometry.\cite{Burkle2012} For the amines
NH$_{2}$-TT1 gives a larger thermopower than NH$_{2}$-TT2. We have
checked that this is a result of the larger negative charge on the
molecule when bonded in $\mbox{NH}_{2}$-TT1 position as compared
to $\mbox{NH}_{2}$-TT2, which moves the HOMO closer to $E_{F}$.
With respect to the thiols we see that both NH$_{2}$-linked geometries
give rise to a thermopower well below those of S-BB and S-TT1 but
still larger than for S-HH. The CN-linked molecules show the largest
$|Q|$. The more positive charge on the molecules in CN-TT1 as compared
to CN-TT2 leads to their smaller, i.e., more negative $Q$. 

Regarding M0 with $\mathrm{X}=\mathrm{S},\mathrm{NH}_{2}$ we can
compare to experimental and theoretical results for $Q$ in the literature.
For biphenyl-diamine a thermopower of $Q_{\mathrm{M0}}^{\mathrm{NH_{2}\textrm{-EXP}}}=4.9\pm1.9$
$\mu$V/K was found at $T=300$ K,\cite{Malen2009a} which compares
reasonably well to our calculated values of $Q_{\mathrm{M0}}^{\mathrm{NH_{2}\textrm{-TT1}}}=10.52\,\mu\mbox{V/K}$
and $Q_{\mathrm{M0}}^{\mathrm{NH_{2}\textrm{-TT2}}}=4.6\,\mu\mbox{V/K}$
for the same $T$. Furthermore, recent calculations within a DFT approach
with an approximate self-interaction correction for comparable geometries
showed similar results to ours.\cite{Quek2011} For biphenyl-dithiol
the comparison is complicated by the fact that our calculated values
vary by two orders of magnitude for the different geometries, i.e.,
$Q_{\mathrm{M0}}^{\textrm{S-HH}}=0.11$ $\mu$V/K, $Q_{\mathrm{M0}}^{\mathrm{\textrm{S-BB}}}=39.14$
$\mu$V/K, $Q_{\mathrm{M0}}^{\mathrm{\textrm{S-TT}}}=28.08$ $\mu$V/K
at $T=300\,\mbox{K}$. They scatter indeed around the experimental
result of $Q_{\mathrm{M0}}^{\mathrm{\textrm{S-EXP}}}=12.9\pm2.2$
$\mu$V/K.\cite{Reddy2007} To our knowledge, the thermopower of cyano-terminated
biphenyls has not yet been reported. A trend by the DFT calculations
to overestimate the thermopower can nevertheless be recognized.\cite{Quek2011}
It is expected from the typical overestimation of experimental conductance
values,\cite{Burkle2012} attributed mostly to the interpretation
of Kohn-Sham eigenvalues as approximate quasi-particle energies.\cite{Quek2007,Strange2011}
According to Eqs.~(\ref{eq:a_2LM}) and (\ref{eq:b_2LM}) an underestimation
of $\left|\tilde{\varepsilon}\right|$ leads to an overestimation
of $|Q|$. However, finite temperature effects due to vibrations,
not accounted for in our calculations, may also play a role in the
room-temperature experiments.

The transport through the well-conjugated molecules M0-M4 is dominated
by the $\pi$ electrons, and we have shown in Refs.~\onlinecite{Mishchenko2010,Mishchenko2011,Burkle2012}
that for these molecules the conductance arises from one transmission
eigenchannel of $\pi$ character.\cite{Mishchenko2011,Burkle2012}
Hence we would expect their thermopower to follow Eq.~(\ref{eq:Q_abcos2}).
Despite the variations of $Q$ with anchoring groups and binding positions,
we find a weak $\mbox{cos}^{2}\varphi$-like decrease of the absolute
values for M1-M4 for all types of geometries. M0, however, deviates
from this trend. Although the electron-donating effect of the alkyl
chains is expected to be small, it increases $Q$ for M1-M4 as compared
to M0. To clarify this we calculated by means of electrostatic potential
fitting and a L\"owdin population analysis the charge transferred
from the alkyl side chains to the two phenyl rings for the hydrogen-terminated
($\mathrm{X}=\mathrm{H}$ in Fig.~\ref{fig:Chemical-structure}),
isolated gas-phase molecules. Both methods yield an overall negative
charge on the phenyl rings which is practically independent of the
alkyl chain length. Therefore the substituent-related energy shift
of frontier orbital levels is similar for M1-M4, and the $a+b\cos^{2}\varphi$
dependence is observed. 

Focusing on the thermopower of M1-M4, we extract $a$ and $b$ by
fitting their $Q$ with Eq.~(\ref{eq:Q_abcos2}). The precise values
are given in Table \ref{tab:Fitting-parameters-obtained}, and the
corresponding fits are shown as continuous lines in Fig.~\ref{fig:QphiDFT}.
Additionally, we list in Table \ref{tab:Fitting-parameters-obtained}
the ratio $r=|Q_{\mbox{M1}}-Q_{\mbox{M4}}|/(Q_{\mbox{M1}}+Q_{\mbox{M4}})$,
quantifying the maximal decrease of $Q$ in that subset of molecules.
We find it to vary between 4\% and 31\%. In detail, we observe the
largest relative change for S-HH followed by CN-TT2. CN-TT1, S-BB,
S-TT1, and $\mbox{NH}_{2}$-TT2 all show a similar $r$, while it
is smallest for $\mbox{NH}_{2}$-TT1.

\begin{table}[t]
\begin{centering}
\begin{tabular}{r||c|c||r@{\extracolsep{0pt}.}l}
 & $a$ ($\mu\mbox{V/K}$) & $b$ ($\mu\mbox{V/K}$) & \multicolumn{2}{c}{$r$ (\%)}\tabularnewline
\hline 
\hline 
S-HH & 0.02 & 0.04 & \multicolumn{2}{c}{31}\tabularnewline
S-BB & 0.86 & 0.43 & \multicolumn{2}{c}{13}\tabularnewline
S-TT1 & 0.93 & 0.37 & \multicolumn{2}{c}{13}\tabularnewline
\hline 
$\mbox{NH}_{2}$-TT1 & 0.37 & 0.06 & \multicolumn{2}{c}{4}\tabularnewline
$\mbox{NH}_{2}$-TT2 & 0.15 & 0.05 & \multicolumn{2}{c}{12}\tabularnewline
\hline 
CN-TT1 & -1.44 & -0.89 & \multicolumn{2}{c}{16}\tabularnewline
 CN-TT2 & -0.81 & -0.65 & \multicolumn{2}{c}{20}\tabularnewline
\end{tabular}
\par\end{centering}

\caption{\label{tab:Fitting-parameters-obtained}Parameters $a$ and $b$ of
Eq.~(\ref{eq:Q_abcos2}) used in Fig.~\ref{fig:QphiDFT} to fit
the DFT results of M1-M4 for each type of junction, and the relative
change $r$ of $Q$ between M1 and M4.}
\end{table}

For M7, Eq.~(\ref{eq:Q_abcos2}) is not expected to hold, because
the transport at $\varphi\simeq90^{\circ}$ is not $\pi$-like but
proceeds through transmission eigenchannels of $\pi$-$\sigma$ character.\cite{Pauly2008,Burkle2012}
Furthermore, M7 shows the largest substituent-related shifting effect
on the biphenyl backbone in our family of molecules due to the electron-donating
nature of the four attached methyl side-groups.\cite{Hansch1991,Suresh1998,Pauly2008}
Its thermopower hence arises from a detailed interplay between the
substituent-related shifting and the large torsion angle, as explained
in Ref.~\onlinecite{Pauly2008}. We find that the absolute values
of $Q$ for M7 are generally lower than predicted by the fits with
Eq.~(\ref{eq:Q_abcos2}). Only for S-TT1 and $\mbox{NH}_{2}$-TT2
the thermopower seems to follow the $a+b\cos^{2}\varphi$ dependence
but this is likely coincidental.

\subsection{Transport analysis using the $\pi$-orbital tight-binding model and
the two-level model\label{sub:Ana-TBM-2LM}}

In order to better understand the differences in the thermopower for
the various anchoring groups and binding positions, we need to examine
the parameters $\tilde{\varepsilon},\tilde{t},\tilde{\Gamma}$ of
the 2LM which determine the thermopower according to Eqs.~(\ref{eq:a_2LM})
and (\ref{eq:b_2LM}). We note that the dominant, angle-independent
term $a$ is a function of $\tilde{\varepsilon},\tilde{\Gamma}$ only.
Thus, to discuss main anchor-group- and binding-site-related variations
of $Q$ for the seven different junction types of Fig.~\ref{fig:geos},
it is sufficient to concentrate on these two parameters. The $\varphi$
dependence of $Q$, however, results from the interference of the
hybridizing pair of phenyl-ring frontier orbital levels, and $b$
hence depends also on $\tilde{t}$.

\begin{figure}[!t]
\begin{centering}
\includegraphics[width=0.7\columnwidth]{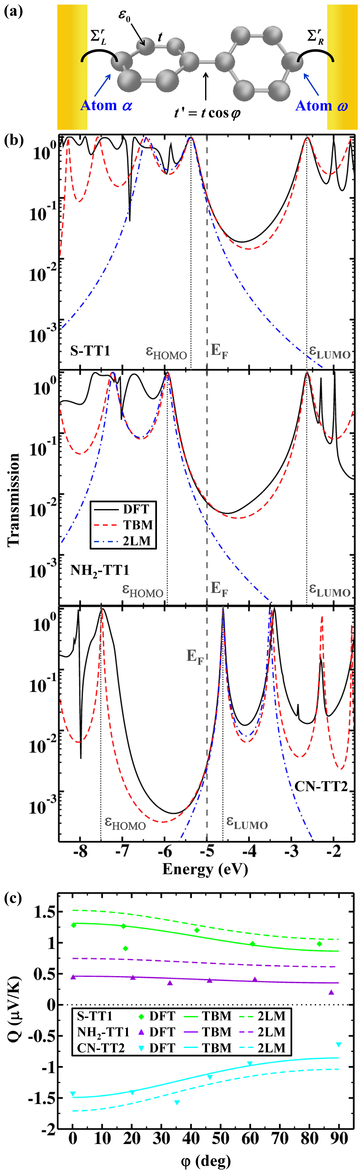}
\par\end{centering}

\centering{}\caption{\label{fig:DFT-TB-2LM}(Color online) (a) Schematic of the TBM used
to fit DFT-based transmission curves. (b) Transmission of M2 as a
function of energy calculated with DFT, and the fits using the TBM
and the 2LM. (c) $Q$ as a function of $\varphi$, comparing values
obtained with the TBM and the 2LM to the DFT-based results. In panels
(b) and (c) S-TT1, NH$_{2}$-TT1, and CN-TT2 junction geometries were
selected.}
\end{figure}

We obtain the parameters of the 2LM from the TBM introduced in Ref.~\onlinecite{Viljas2008}.
The TBM is sketched in Fig.~\ref{fig:DFT-TB-2LM}(a). Similar to
the 2LM, the H\"uckel-like TBM is characterized by three parameters
which are the on-site energy $\varepsilon_{0}$ of each carbon atom,
the nearest-neighbor hopping $t$ between atoms on each of the phenyl
rings, and the electrode-related broadening $\Gamma$. The inter-ring
hopping is given as $t'=t\cos\varphi$. Using the wide-band approximation,
we assume all components of the lead self-energy matrices to vanish
except for $\left(\Sigma_{L}^{r}\right)_{\alpha\alpha}=\left(\Sigma_{R}^{r}\right)_{\omega\omega}=-\mbox{i}\mbox{\ensuremath{\Gamma}/2}$,
with $\alpha$ and $\omega$ indicating the terminal carbon atoms
of the biphenyl molecule as shown in Fig.~\ref{fig:DFT-TB-2LM}(a). 

The parameters $\varepsilon_{0},t,\Gamma$ of the TBM are extracted
by fitting $\tau(E)$ curves calculated with DFT. We focus on the
molecules M1-M4 and set $\varphi$ to the torsion angle realized in
the specific junction geometry (see Table \ref{tab:phiQtable-DFT}).
Concentrating particularly on the HOMO-LUMO gap and frontier orbital
peaks, we find that the fitted TBM generally reproduces well the transmission
in that range and that the parameters extracted for M1-M4 are very
similar in each of the seven types of junctions. Finally, the parameters
of the 2LM are derived from those of the TBM. $\tilde{\varepsilon}$
and $\tilde{t}$ are obtained by evaluating appropriate eigenvalues
of the angle-dependent H\"uckel-like Hamiltonian of the TBM. For
$\tilde{\Gamma}$ we identify imaginary parts of complex eigenvalues
of the non-Hermitian matrices $(H+\Sigma_{L}^{r}+\Sigma_{R}^{r})_{jk}$
for the TBM and the 2LM, respectively.\cite{Burkle2012} Here, $H_{jk}$
and $(\Sigma_{L}^{r})_{jk},(\Sigma_{R}^{r})_{jk}$ represent the matrix
elements of the Hamiltonian and of the electrode self-energies in
the corresponding model. All the parameters determined in this way
are listed in Table \ref{tab:TBparams}. For M2, transmission curves
calculated with the DFT, the TBM, and the 2LM are shown in Fig.~\ref{fig:DFT-TB-2LM}(b)
for each of the three anchoring groups.

\begin{table}[tb]
\begin{centering}
\begin{tabular}{c||c|c|c||c|c|c}
 & $\varepsilon_{0}$ & $t$ & $\Gamma$ & $\tilde{\varepsilon}$ & $\tilde{t}$ & $\tilde{\Gamma}$\tabularnewline
\hline 
\hline 
S-HH & -4.40 & -2.30 & 0.70 & -1.70 & -0.68 & 0.22\tabularnewline
S-BB & -4.02 & -1.95 & 1.10 & -0.97 & -0.58 & 0.35\tabularnewline
S-TT1 & -4.00 & -1.90 & 0.96 & -0.90 & -0.56 & 0.31\tabularnewline
\hline 
$\mbox{NH}_{2}$-TT1 & -4.30 & -2.29 & 0.60 & -1.59 & -0.68 & 0.19\tabularnewline
$\mbox{NH}_{2}$-TT2 & -4.40 & -2.32 & 0.66 & -1.72 & -0.69 & 0.21\tabularnewline
\hline 
CN-TT1 & -6.10 & -2.00 & 0.14 & 0.90 & -0.59 & 0.04\tabularnewline
CN-TT2 & -6.05 & -1.99 & 0.15 & 0.94 & -0.59 & 0.05\tabularnewline
\end{tabular}
\par\end{centering}

\caption{\label{tab:TBparams}Parameters of the TBM $\varepsilon_{0}$, $t$,
$\Gamma$ obtained by fitting the DFT-based transmission curves for
M1-M4 for each type of junction. The parameters $\tilde{\varepsilon}$,
$\tilde{t}$, $\tilde{\Gamma}$ of the 2LM are derived from those
of the TBM as described in the text. All values are given in units
of eV.}
\end{table}

Using the parameters of Table \ref{tab:TBparams} we compare in Fig.~\ref{fig:DFT-TB-2LM}(c)
$Q$ as a function of $\varphi$ for the TBM and 2LM fits with the
DFT results. We find that the TBM agrees well with the DFT-based values
for the illustrated junction geometries S-TT1, NH$_{2}$-TT1, and
CN-TT2. The 2LM, instead, overestimates $|Q|$ somewhat. Considering
Eq.~(\ref{eq:QlowT}) and the transmission curves in Fig.~\ref{fig:DFT-TB-2LM}(b),
we attribute this to an underestimation of $\tau(E_{F})$ and an overestimation
of $|\partial_{E}\tau(E_{F})|$. All results exhibit a consistent
weak dependence of $Q$ on $\varphi$.

The data in Table \ref{tab:TBparams} shows that transport through
the biphenyl molecules is off-resonant with the relation $\tilde{\Gamma}\ll\left|\tilde{\varepsilon}\right|$
being well fulfilled. As argued in Sec.~\ref{sub:Qphi-Models}, $a,b$
should thus take the same sign and change it together with $Q$ when
the transport for S- and NH$_{2}$-linked molecules changes from HOMO-
to LUMO-dominated for CN anchors. This is consistent with our findings
in Figs.~\ref{fig:QphiDFT} and \ref{fig:DFT-TB-2LM}(c), and explains
the decrease of $|Q|$ with increasing $\varphi$.%
\footnote{We note that Fig.~2(a) of Ref.~\onlinecite{Finch2009} suggests
that transport is strongly off-resonant also in their calculations,
and the positive $Q$ should hence decrease with $\varphi$ in their
Fig.~3(b).%
}

Coming back to our discussion of the differences of the thermopower
for the various anchoring groups and binding positions in Fig.~\ref{fig:QphiDFT},
we observe that $\tilde{\varepsilon}$ is around 0.6 to 0.8 eV closer
to $E_{F}$ for S-BB and S-TT1 as compared to S-HH, $\mbox{NH}_{2}$-TT1,
and $\mbox{NH}_{2}$-TT2, which explains their larger $Q$. For the
CN-terminated molecules $|\tilde{\varepsilon}|$ is comparable to
those for S-BB and S-TT1. Slightly larger values of $|Q|$ for CN
result from the very small broadenings $\tilde{\Gamma}$. Furthermore,
for both $\mbox{NH}_{2}$ and CN, $\tilde{t}$ and $\tilde{\Gamma}$
are essentially independent of the binding position, and the difference
in $Q$ between TT1 and TT2 hence stems from the changes in the alignment
of the HOMO and LUMO levels.

\section{Conclusions\label{sec:Conclusions}}

We have analyzed theoretically the thermopower of single-molecule
junctions consisting of biphenyl derivatives contacted to gold electrodes.
Or DFT-based study with the three anchors S, NH$_{2}$, and CN shows
a positive thermopower for S or $\mbox{NH}_{2}$ and a negative one
for CN. For the junction geometries considered, different binding
sites did not affect the sign of $Q$ but led to variations of absolute
value. For thiolated molecules in bridge and top binding sites $Q$
can be up to two orders of magnitude larger than for molecules bonded
in hollow position, while the variations for the two considered top
binding sites were around a factor of two for $\mbox{NH}_{2}$ and
CN anchors. We have explained these observations by the changes in
the level alignment of current-carrying frontier molecular orbitals.
They are caused by the binding-site-dependent charge transfer at the
metal-molecule interface.

The main purpose of this work was the study of the dependence of the
thermopower on conjugation for an experimentally relevant system.
In our set of six biphenyl derivatives, the conjugation was controlled
by the torsion angle $\varphi$ between the phenyl ring planes, and
it was varied stepwise between $0$ and $90^{\circ}$ by means of
alkyl side chains attached to the molecules. Despite the sensitivity
of the thermopower to the precise geometry at the molecule-metal interface,
we observed for all investigated types of junction configurations
a decrease in $|Q|$ with increasing $\varphi$, following a characteristic
$a+b\cos^{2}\varphi$ law. We explained this behavior in terms of
a two-level model, which considers the pair of hybridizing frontier
orbitals on the phenyl rings. Predictions by this model of a simultaneous
change in sign of $Q,a,b$ for a change from HOMO- to LUMO-dominated
transport in the off-resonant situation are consistent with our DFT
results. Overall, the influence of conjugation on the thermopower
is much less pronounced than on the conductance.

We propose to measure the $a+b\cos^{2}\varphi$ dependence of the
thermopower for the set of biphenyl molecules studied here. Using
alkyl chains of different lengths, parasitic substituent-related shifts
in $Q$, superimposed on the weak $a+b\cos^{2}\varphi$ dependence,
are largely avoided. Depending on binding site and employed anchoring
group, relative variations of $Q$ of around 15\% are expected between
M1 and M4. Since frontier molecular orbital energies are likely positioned
closer to $E_{F}$ in our calculations than in the experiment, the
relative changes of $Q$ with $\varphi$ are expected to be somewhat
smaller than in our theoretical predictions. Nevertheless, we suggest
that the variations of the thermopower with torsion angle are experimentally
detectable.
\begin{acknowledgments}
We acknowledge fruitful discussions with A.\ Bagrets, F.\ Evers,
and V.\ Meded. R.\ Ahlrichs and M.\ Sierka are thanked for providing
us with TURBOMOLE. M.B.\ and G.S.\ were supported through the DFG
Center for Functional Nanostructures (Project C3.6), the DFG priority
program 1243, and the Initial Training Network ``NanoCTM'' (Grant
No.\ FP7-PEOPLE-ITN-2008-234970), F.P.\ through the Young Investigator
Group, L.A.Z. by the EU through the BIMORE Network (MRTN-CT-2006-035859),
and J.K.V.\ through the Academy of Finland. D.V.\ and M.M.\ acknowledge
funding by the Swiss National Science Foundation and the Swiss National
Center of Competence in Research ``Nanoscale Science''. The work
of A.M.\ and T.W.\ was financed by the Swiss National Science Foundation
(200021.124643, NFP62), the Initial Training Network FUNMOLS, the
DFG priority program 1243, and the University of Bern.
\end{acknowledgments}

\end{document}